\documentclass[12pt,twoside]{article}
\usepackage{amsmath}

\begin{document}

\title{Noncommuting Coordinates in the Landau problem}
\author{Gabrielle Magro\\
\small MIT Department of Physics\\ 
\small Senior thesis, submitted on 17 January 2003} \date{}
\maketitle

\pagestyle{myheadings}
\markboth{G. Magro}{Noncommuting Coordinates in the Landau Problem}
\thispagestyle{empty}

\begin{abstract}
\noindent
Basic ideas about noncommuting coordinates are summarized, and then 
coordinate noncommutativity, as it arises in the Landau problem, is 
investigated. 
I review a quantum solution to the Landau problem, and evaluate the
coordinate commutator in a truncated state space of Landau levels.
Restriction to the lowest Landau level reproduces the well known 
commutator of planar coordinates. Inclusion of a finite number of 
Landau levels yields a matrix generalization. 
\end{abstract}

\section{Introduction}

Because of its relevance to string theory, the idea of noncommuting 
spatial coordinates
has gained much attention recently, but the idea actually predates
string 
theory. 
Coordinate noncommutativity, defined by the equation
\begin{equation}
[x^i ,x^j ]=i\theta ^{ij}
\end{equation}	
where $\theta ^{ij}$ is a constant, anti-symmetric two-index object, 
implies
a coordinate uncertainty relation, resulting in a discretization of
space 
itself (similar
to the familiar quantum phase space) and the elimination of spatial 
singularities. 
The idea was suggested by Heisenberg in the 1930s 
as a way to eliminate divergences in quantum field theory arising from
the 
assumption of point interactions between fields and matter, and the
first 
paper
on the subject appeared some years later [1]. Since 
then, much attention has been given to the study of quantum field theory 
on noncommutative spaces. Because quantum mechanics is just the 
one-particle nonrelativistic sector 
of quantum field theory, it is also relevant to study the quantum 
mechanics of particles in noncommutative spaces, and to understand the 
transition between the commutative and noncommutative
regimes. 

There exists a well known phenomenological realization of 
noncommuting coordinates
in the realm of quantum mechanics: a charged particle in an external 
magnetic field so strong that projection to the lowest Landau level is 
justified.A charged particle in an external magnetic field is
effectively 
confined to 
a two-dimensional space perpendicular to the field, which becomes
noncommuting when motion is projected onto the lowest Landau level. 
It is interesting to note that this example is similar to the instances
of 
noncommuting coordinates which arise in string theory in that both are 
characterized by the 
presence of a strong background magnetic-like field.  
Because the lowest Landau level is the only physically realized example
of 
noncommuting coordinates, 
it should be useful to our understanding of noncommutative spaces in 
physics to know
precisely how noncommutativity arises in this system.  To this end, 
a quantum solution to the Landau problem is
reviewed, and then the coordinate commutator is calculated after a 
projection 
to a truncated space of Landau levels. 

\section{The Landau Problem}
We consider a charged ($e$) and massive ($m$) particle in a constant 
magnetic
field ($B$), which we chose to point along the $z$ axis without loss 
of generality. This is called the Landau problem because
the eigenstates and eigenvalues were investigated for the first time by 
Landau. 
The Lagrangian describing the particle's motion is
\begin{equation}
L=\frac{1}{2}m(\dot{x}^2+\dot{y}^2+\dot{z}^2) 
+\frac{e}{c}(\dot{x}A_{x}+\dot{y}A_{y}+\dot{z}A_{z}).
\end{equation}
>From this, we determine the Hamiltonian for the particle in the usual
way, 
with the result
\begin{equation}
H=\frac{1}{2m}(\vec{p}-\frac{e}{c}\vec{A})^2
\end{equation}
where the canonical momentum $\vec{p}$ of the particle is no longer the 
usual
$m\vec{v}$ but is equal to $m\vec{v} + \frac{e}{c}\vec{A}$. The 
Hamiltonian is thus simply
$H=\frac{1}{2}m\vec{v}^2$, which is what we expect since the magnetic 
field 
does no work and thus cannot contribute to the energy.

To solve the eigenstates and eigenenergies of this problem, we must 
prescribe
a vector potential $\vec{A}$, satisfying
$\nabla \times \vec{A}=B\hat{z}$. We chose $\vec{A}=(0,xB,0)$
Substituting this gauge choice into the Hamiltonian, we have
\begin{equation}
H=\frac{1}{2m} (p_x ^2 + p_z ^2 + (p_y - \frac{exB}{c} )^2 ).
\end{equation}
Because $H$ commutes with both $p_z$ and with $p_y$, we can write 
the eigenvalue equations $p_z|\Psi \rangle = \hbar k_z |\Psi \rangle$
and $p_y|\Psi \rangle = \hbar k_y |\Psi \rangle$, 
where $|\Psi \rangle$ denotes the eigenstates of the full Hamiltonian,
and 
$\hbar k_z$ and $\hbar k_y$ are the eigenvalues of their respective 
momentum operators. 
Using these eigenvalues, the Hamiltonian can be written as
\begin{equation}
H=\frac{1}{2m}(\hbar k_z) ^2+\frac{1}{2m} \left (p_x ^2 
+\left (\frac{eB}{c} \right )^2 \left (x-\frac{c\hbar k_y}{eB} \right
)^2  
\right ).
\end{equation}
>From now on, motion in the $z$ direction is suppressed since it is not 
quantized, and
only planar motion in the $x$-$y$ plane is retained. The second term in 
the 
Hamiltonian is just a shifted harmonic oscillator with angular frequency 
$\omega = \frac{eB}{cm}$. 
The eigenstates $|\Psi \rangle$ are thus labeled by the number of 
oscillator quanta $n$ and by 
$k_y$. The associated eigenenergies are
\begin{equation}
E_n = \hbar \frac{eB}{mc} \left (n+\frac{1}{2} \right ).
\end{equation}
$k_y$ is free, so each $n$ indexes an energy eigenstate, called a Landau 
level, which is 
infinitely degeneracy in $k_y$. Note that adjacent Landau levels are 
separated by 
energy $\hbar \frac{eB}{mc}$. 
The eigenfuctions in coordinate space are
\begin{equation}
\langle x,y| n,k_y \rangle = \frac{1}{\sqrt{2\pi \hbar}}e^{ik_y y}\phi
_n 
(x-\frac{c\hbar k_y}{eB})
\end{equation}
where $\phi _n$ are the normalized harmonic oscillator wavefunctions. 
>From now on, we let $k\equiv \hbar k_y$ and label the eigenstates
$|n,k\rangle$.

\section{Projection to the lowest Landau level}
Since the separation between the states $|n,k\rangle $ is 
$\mathcal{O}(B/m)$, if the magnetic
field is strong, only the lowest Landau level $|0,k\rangle $ is
relevant. 
The higher states
are essentially decoupled to infinity.  The large $B$ limit is
equivalent 
to 
the small $m$ limit. 
So the Lagrangian can be modified to describe only the lowest Landau
level 
by setting $m$ to 
zero in (2). With this modification, and the addition of a
potential $V(x,y)$
to represent impurities in the plane, the Lagrangian becomes
\begin{equation}
L_{lLl}=\frac{e}{c}Bx\dot{y}-V(x,y).
\end{equation}
This has the same form as $L=p\dot{q}-H(p,q)$, and thus we recognize 
$\frac{e}{c}Bx$ and $y$
as canonical conjugates, with the corresponding commutator: 
\begin{eqnarray}
[\frac{eB}{c}x,y] &=& -i\hbar \\
\rightarrow [x,y]&=&-i\frac{\hbar c}{eB}
\end{eqnarray}
So we see that by restricting the particle to the first Landau level,
the 
space 
it moves in no longer 
obeys the standard Heisenberg algebra. This is called the ``Peierls 
substitution" [2].

We can verify this result in a less heuristic fashion, by directly 
calculating the matrix elements of 
the coordinate commutator [3]. 
\begin{eqnarray}
\langle n, k |[x,y]|n',k'\rangle &=& \langle n,k |xy |n',k'  \rangle - 
\langle 
n,k |yx  |n',k' \rangle \\
&=& \langle n,k | xy | n', k' \rangle -  \langle n', k'  | yx | n, k 
\rangle ^* 
\\
&=& f(nk,n'k')-f^*(n'k',nk)
\end{eqnarray}
where
\begin{equation}
f(nk,n'k')=\langle n,k |xy  |n',k' \rangle 
\end{equation}
We evaluate $f$ by inserting a complete set of states in the 
product $xy$:
\begin{equation}
f(nk,n'k')=  \sum _{m,q} \langle n,k| x |m,q \rangle \langle m,q | y 
|n',k' \rangle.
\end{equation}
If we restrict the particle's world to include only the lowest Landau 
level, 
then we should only include the lowest Landau level (with its infinite 
degeneracy) 
in the intermediate state sum of this calculation. Additionally, since 
we're pretending
that the world is confined to the lowest Landau level, it only makes
sense 
to calculate
this matrix element for $n=n'=0$. To evaluate that element, we must 
calculate
\begin{equation}
f(0k,0k')= \int dq \langle 0,k | x | 0,q \rangle \langle 0,q | y | 0, k' 
\rangle.
\end{equation}
To evaluate each matrix element, we explicitly write the
plane wave functions and the harmonic oscillator wavefunctions. These
expressions are readily evaluated by integration over the $x$-$y$ plane. 
\begin{eqnarray}
\langle 0,k | x | 0,q \rangle &=& \int dx dy \frac{1}{\sqrt{2\pi 
\hbar}}e^{-iky/ \hbar}
(\frac{m \omega}{\pi \hbar})^{1/4} 
e^{-\frac{m\omega}{2\hbar}(x-\frac{ck}{eB})^2}x \nonumber \\
 && \frac{1}{\sqrt{2\pi \hbar}}e^{iqy/ \hbar}
(\frac{m \omega}{\pi \hbar})^{1/4} 
e^{-\frac{m\omega}{2\hbar}(x-\frac{cq}{eB})^2} \\
&=& \delta (k-q)\frac{cq}{eB} \\
\langle 0,q | y | 0, k' \rangle &=& \int dx dy \frac{1}{\sqrt{2\pi 
\hbar}}e^{-iqy/ \hbar}
(\frac{m \omega}{\pi \hbar})^{1/4} 
e^{-\frac{m\omega}{2\hbar}(x-\frac{cq}{eB})^2}y \nonumber \\
&& \frac{1}{\sqrt{2\pi \hbar}}e^{ik'y/ \hbar}
(\frac{m \omega}{\pi \hbar})^{1/4} 
e^{-\frac{m\omega}{2\hbar}(x-\frac{ck'}{eB})^2} \\
&=& i\hbar \delta'(q-k')
\end{eqnarray}
where the prime denotes differentiation with respect to the argument. 

The commutator can then be obtained:
\begin{equation}
\langle 0,k|[x,y]  |0,k' \rangle=-i\frac{\hbar c}{eB} \langle 0,k|0,k'  
\rangle
\end{equation}
which is consistent with (10).

\section{Projection to higher Landau levels}
One could now pretend that the world is restricted to the lowest $N +1$ 
Landau levels ($n=0,1,...,N$).
In this case, we cannot obtain the commutator $[x,y]$ heuristically by 
modifying the Lagrangian and
reading off a canonical pair, but our method of explicitly calculating
the 
relevant matrix 
elements is still valid. To evaluate $f$ we include the lowest $N+1$ 
levels 
(each with its infinite
degeneracy) in the intermediate state sum.
\begin{equation}
f(nk,n'k')=\sum _{m=0} ^{N} \int dq \langle n,k | x | m,q \rangle
\langle 
m,q | y | n', k' \rangle.
\end{equation}
This is defined for $n$ and $n'$ less than or equal to $N$.
Each matrix element is evaluated by explicitly writing the coordinate 
space plane 
wavefunctions, and representing the harmonic oscillator wavefunctions as 
$\phi _n (x)$.
The expressions are  simplified by exploiting
the orthonormality of the harmonic oscillator wavefunctions and
integrating over the plane.
\begin{eqnarray}
\langle n,k | x | m,q \rangle &=& \int dx dy \frac{1}{\sqrt{2\pi \hbar}}
e^{-iky/ \hbar}\phi _n(x-\frac{ck}{eB})x \nonumber \\
 && \frac{1}{\sqrt{2\pi \hbar}} e^{iqy/\hbar}\phi _m (x-\frac{cq}{eB}) \\
 &=&\delta (q-k)\delta _{nm} \frac{ck}{eB} + \delta (q-k)\langle n|x|m 
\rangle \\
\langle m,q | y | n', k' \rangle &=& \int dx dy \frac{1}{\sqrt{2\pi 
\hbar}}
e^{-iqy/ \hbar}\phi _n(x-\frac{cq}{eB})y \nonumber \\
&& \frac{1}{\sqrt{2\pi \hbar}} e^{ik'y/\hbar}\phi _{n'}
(x-\frac{ck'}{eB}) 
\\
&=& \hbar i \delta '(q-k')\delta _{mn'}-\frac{c}{eB}\delta (q-k')\langle 
n' |p|m\rangle
\end{eqnarray}

By evaluating the sums and integrals of (22), one can find that
the matrix elements of the coordinate commutator vanish unless $n=n'=N$. 
For this
case, we get the result
\begin{equation}
\langle N,k | [x, y] | N,k' \rangle = -i\hbar \frac{c}{eB}(N+1)\delta 
(k-k').
\end{equation}
So, for example, the matrix representation of the coordinate commutator 
for a particle confined to the two lowest Landau levels is
\begin{equation}
[x,y]= \left ( \begin{array}{cc} 0 & 0 \\ 0 & -2i\hbar \frac{c}{eB}
\end{array} \right)
\end{equation}
If we expanded our world to include another Landau level, we would find 
that
the matrix element $\langle 2,k | [x,y] | 2,k' \rangle$ vanishes because 
of the additional 
$m=2$ term in the sum in (22).  
Thus we have shown that the
phenomena of noncommutative space is not specific to the lowest
Landau level but can be obtained by 
projecting to an arbitrary finite number of Landau levels. [4]

\section{Motion in the symmetric gauge}
By choosing a different gauge, we can verify this result, as well as
come 
upon it much more 
elegantly. 
Substituting the symmetric gauge $\vec{A}=\frac{B}{2}(-y,x)$ into the 
Hamiltonian describing the planar motion, we find
\begin{eqnarray}
H&=&\frac{1}{2m}\left ((p_x +\frac{eB}{2c}y)^2+
(p_y - \frac{eB}{2c}x )^2 
\right ) \\
&=&\frac{1}{2m}(p_x ^2 + p_y ^2) + \frac{1}{2}m 
\left (\frac{eB}{2mc} \right )^2(x^2 + y^2) - \frac{eB}{2mc}L, 
\end{eqnarray}
where $L = xp_y - yp_x$ is the angular momentum in the $x$-$y$ plane. In 
this 
gauge, the
system looks like a two-dimensional harmonic oscillator with 
an additional 
interaction $-\frac{eB}{2mc}L$. Since we are dealing here only with 
harmonic 
oscillator
wavefunctions and not plane waves, it is clear that this problem 
will most easily 
solved by introducing harmonic oscillator creation and annihilation 
operators.
To this end, we define:
\begin{eqnarray}
a &=& 
\frac{1}{2}\sqrt{\frac{eB}{2c\hbar}}(x-iy)+\frac{i}{2}\sqrt{\frac{2c}{eB\hbar}}(p_x 
- ip_y) \\
b &=& 
\frac{1}{2}\sqrt{\frac{eB}{2c\hbar}}(x+iy)+\frac{i}{2}\sqrt{\frac{2c}{eB\hbar}}(p_x 
+ ip_y).
\end{eqnarray}
satisfying the following commutation relations
\begin{align}
[a,a^{\dagger}] = [b,b^{\dagger}]=1
\end{align}
with all other commutators vanishing. 
($a, a^{\dagger}$) and ($b,b^{\dagger}$) are two pairs of independent 
harmonic oscillator operators.
We can now write $L$ and $H$ in terms of these operators:
\begin{eqnarray}
L &=& \hbar (a^{\dagger}a-b^{\dagger}b) \\
H &=& \hbar \frac{eB}{2mc}(a^{\dagger}a+b^{\dagger}b +1)-\hbar 
\frac{eB}{2mc}(a^{\dagger}a-b^{\dagger}b) \\
&=& \hbar \frac{eB}{mc}(b^{\dagger}b+\frac{1}{2}).
\end{eqnarray}
Eigenstates of the Hamiltonian are labeled by the number $j$ of excitation
quanta of 
the oscillator $a$, and the number 
$n$ of excitation quanta of the oscillator $b$:  
\begin{eqnarray}
a^{\dagger}a| n,j \rangle = j | n,j \rangle, \\
b^{\dagger}b| n,j \rangle = n | n,j \rangle.
\end{eqnarray}
Both $j$ and $n$ can take on any non-negative integer value. 
Only the $n$ quanta contribute to the energy, so the Landau levels, 
labeled by $n$, are each
infinitely degenerate in $j$. 

To evaluate the commutator $[x,y]$ using this basis, we first rewrite
$x$ 
and 
$y$ in terms of $\alpha$ and $\alpha ^{\dagger}$, where
$\alpha = a + b^{\dagger}$
\begin{eqnarray}
x&=&\sqrt{\frac{ \hbar c}{2eB}}(\alpha + \alpha ^{\dagger}), \\
y&=& i\sqrt{\frac{ \hbar c}{2eB}}(\alpha - \alpha ^{\dagger}).
\end{eqnarray}
Then the coordinate commutator is
\begin{eqnarray}
[x,y]&=&\frac{i}{2}\frac{\hbar c}{eB}[\alpha + \alpha ^{\dagger}, \alpha
- 
\alpha ^{\dagger} ] \\
&=& -i\frac{\hbar c}{eB}[\alpha, \alpha ^{\dagger}].
\end{eqnarray}
It we evaluate this exactly, we find that the commutator vanishes as 
expected.

Now using this new basis we can explicitly evaluate matrix elements
for an incomplete state space. We 
will
denote the eigenstates  $| n,j \rangle$ explicitly as product states: 
$|n\rangle |j \rangle$. To determine the commutator, we need to 
evaluate the matrix elements:
\begin{equation}
\langle n | \langle j |\alpha \alpha ^{\dagger} | j' \rangle | n'
\rangle 
- 
\langle n | \langle j |\alpha ^{\dagger} \alpha |j' \rangle |n' \rangle 
\equiv (1) - (2).
\end{equation}
We evaluate each term separately by inserting intermediate states. 
\begin{eqnarray}
(1) &=& \sum _{m, l = 0}^{m=N, l= \infty} \langle n | \langle j | \alpha
|l \rangle |m \rangle
\langle m | \langle l |\alpha ^{\dagger}|j' \rangle |n' \rangle \\ 
(2)&=&\sum _{m,l=0} ^{m=N , l=\infty} \langle n| 
\langle j|\alpha ^{\dagger} |l 
\rangle |m \rangle
 \langle m| \langle l|\alpha | j' \rangle | n' \rangle
\end{eqnarray}
Summing over $l$, one readily finds
\begin{eqnarray}
(1)-(2)= \delta _{nn'} \delta _{jj'} ( 1+ \sum _{m=1} ^{N +1} m \delta 
_{nm} 
-\sum _{m=0}^{N - 1} (m +1) \delta _{nm}   )
\end{eqnarray}
Since $m$ never gets larger than $N$, we can write this as
\begin{eqnarray}
&=& \delta _{nn'} \delta _{jj'} (1+\sum _{m=1}^{N}m \delta _{nm} - \sum 
_{m=0}^{N - 1}(m+1)\delta _{nm}) \\
&=&\delta _{nn'} \delta _{jj'}(1+N \delta _{nN} - \sum 
_{m=0} ^{N - 1}\delta _{nm}) 
\end{eqnarray}
Let us analyze this expression. It clearly vanishes for off diagonal
elements. For $n=n'\neq  N$, it also vanishes because the sum will yield
unity since $n$ is necessarily less than $N$.  For $n=n'=N$, 
we obtain our earlier result: 
\begin{eqnarray}
\langle N | \langle j | [x,y] | j' \rangle | N \rangle = -i\frac{\hbar 
c}{eB}
(N + 1)\langle j|j' \rangle.
\end{eqnarray}

Suppose we had let $N \rightarrow \infty$, then (46) would have been
\begin{eqnarray}
&=&\delta _{nn'} \delta _{jj'}(1+\sum _{m=1}^{\infty }m\delta _{nm} -
\sum 
_{m=0} ^{\infty }(m+1)\delta _{nm}) \\
&=& \delta _{nn'} \delta _{jj'} (1 - \delta _{n0}-\sum _{m=1} ^{\infty }
\delta _{nm})
\end{eqnarray}
which clearly vanishes for any choice of $n$. 
Noncommutativity in the Landau problem is clearly associated with 
a truncated state space.

\section{Acknowledgements}
Thanks to Roman Jackiw for giving me this problem to solve.

\end{document}